\documentclass[preprint,showpacs,showkewords,preprintnumber,amsmath,amssymb]{revtex4}
 \usepackage{tipa}
 \usepackage{amsmath}
 \usepackage{txfonts}
 \usepackage{amssymb}
 \usepackage{graphicx,epsfig}
 \usepackage{bm}
 \begin{document}

 \title{ New partial symmetries from group algebras for lepton mixing }
 \author{Shu-Jun Rong}\email{rongshj@glut.edu.cn}

 \affiliation{College of Science, Guilin University of Technology, Guilin, Guangxi 541004, China}

 \begin{abstract}
Recent stringent experiment data of neutrino oscillations induces partial symmetries such as $Z_{2}$, $Z_{2}\times CP$ to derive lepton mixing patterns. New partial symmetries expressed with elements of group algebras are studied. A specific lepton mixing pattern could correspond to a set of equivalent elements of a group algebra. The transformation
which interchanges the elements could express a residual $CP$ symmetry. Lepton mixing matrices from  $S_{3}$ group algebras are of the trimaximal form with the $\mu-\tau$ reflection symmetry. Accordingly, elements of $S_{3}$ group algebras are equivalent to $Z_{2}\times CP$. Comments on $S_{4}$ group algebras are given. The predictions of $Z_{2}\times CP$  broken from the group $S_{4}$ with the generalized $CP$ symmetry are also obtained from elements of $S_{4}$ group algebras.

\end{abstract}


 \maketitle
 \section{Introduction}
Discoveries of neutrino oscillation\cite{1,2,3} opened a window to physics beyond standard model. In order to explain possible patterns of lepton mixing parameters, discrete flavor symmetries were extensively investigated in recent decades\cite{4,5,6,7,8,9,10,11,12,13,14,15,16,17,18,19,20,21,55,56,57,58,59,60,61,62,63}. The general route on this approach is as follows. First suppose that the Lagrangian of leptons is invariant under actions of some finite group $G_{f}$. After symmetry breaking from vacuum expectation values of scalar multiplets, $G_{f}$ is reduced to $G_{e}$ in the charged lepton section and $G_{\nu}$ in the neutrino section respectively. Accordingly, the mass matrix of charged leptons is invariant under some unitary transformation, i.e.,
\begin{equation}
\label{eq:1}
X_{e}^{+}M_{e}^{+}M_{e}X_{e}=M_{e}^{+}M_{e}.
\end{equation}
So we have
\begin{equation}
\label{eq:2}
U_{e}^{+}M_{e}^{+}M_{e}U_{e}=diag(m_{e}^{2},~~m_{\mu}^{2},~~ m_{\tau}^{2}),
\end{equation}
\begin{equation}
\label{eq:3}
U_{e}^{+}X_{e}U_{e}=diag(e^{i\alpha_{1}},~e^{i\alpha_{2}},~~e^{i\alpha_{3}}).
\end{equation}
The counterparts for Dirac neutrinos are written as
\begin{equation}
\label{eq:4}
X_{\nu}^{+}M_{\nu}^{+}M_{\nu}X_{\nu}=M_{\nu}^{+}M_{\nu},
\end{equation}
\begin{equation}
\label{eq:5}
U_{\nu}^{+}M_{\nu}^{+}M_{\nu}U_{\nu}=diag(m_{1}^{2},~~m_{2}^{2},~~ m_{3}^{2}),
\end{equation}
\begin{equation}
\label{eq:6}
U_{\nu}^{+}X_{\nu}U_{\nu}=diag(e^{i\beta_{1}},~e^{i\beta_{2}},~~e^{i\beta_{3}}).
\end{equation}
For Majorana neutrinos, they read
\begin{equation}
\label{eq:7}
X_{\nu}^{T}M_{\nu}X_{\nu}=M_{\nu},
\end{equation}
\begin{equation}
\label{eq:8}
U_{\nu}^{T}M_{\nu}U_{\nu}=diag(m_{1},~~m_{2},~~ m_{3}),
\end{equation}
\begin{equation}
\label{eq:9}
U_{\nu}^{+}X_{\nu}U_{\nu}=diag(\pm1,~\pm1,~~\pm1).
\end{equation}
So residual symmetries $X_{e}$, $X_{\nu}$ can determine the lepton mixing matrix $U_{PMNS}\equiv U_{e}^{+}U_{\nu}$ up to permutations of rows or columns.

However, mixing patterns based on small flavor groups cannot accommodate new stringent experiment data, especially the nonzero mixing angle $\theta_{13}$. Although some large groups could give a viable $\theta_{13}$, the Dirac $CP$ violating phase from them is trivial\cite{22}. In order to alleviate the tension between predictions of flavor groups and experiment constraints, one can resort to partial symmetries. Namely, the lepton mixing matrix is partially determined by symmetries such as  $Z_{2}$\cite{23,24,25}, $Z_{2}\times CP$\cite{26,27,28,29,30,31,32,33,34,35,36,37,38,39,40,41,42,43}. Here $CP$ denotes a generalised $CP$ transformation(GCP). For $Z_{2}$ symmetries, an unfixed unitary rotation is contained in the mixing matrix. Even so, they may predict some mixing angle, Dirac CP phase or correlation of them. If the residual symmetry is ($Z_{2e}\times Z_{2e}$, $Z_{2\nu}\times CP_{\nu}$) or ($Z_{ne}$, $Z_{2\nu}\times CP_{\nu}$) with $n\geq3$, the Dirac $CP$ phase would be trivial or maximal in the case that the residual flavor group is from small groups $S_{4}$, $A_{5}$\cite{30,32,39}. Here the symmetries of the charged lepton sector and those of neutrinos are marked with the subscripts $e$, $\nu$ respectively. To obtain a more general $CP$ phase, one can choose the residual symmetry ($Z_{2e}\times CP_{e}$, $Z_{2\nu}\times CP_{\nu}$)\cite{44,45}. Then the lepton mixing matrix contains two angle parameters to constrain by experiment data.

In this paper we explore a new construct to describe partial symmetries which was proposed recently in Ref.~\cite{46}. The partial symmetry is expressed by an element of a group algebra. According to the Ref.~\cite{53}, a group algebra $K[G]$ is the set of  all linear combinations of elements of the group $G$ with coefficients in the field $K$. A general element of $K[G]$ is denoted as
\begin{equation}
\label{eq:10}
 \sum_{g\in G} a_{g}g .
\end{equation}
$K[G]$ is an algebra over $K$ with the addition and multiplication defined respectively as
\begin{equation}
\label{eq:11}
 \sum_{g\in G} a_{g}g+\sum_{g\in G} b_{g}g=\sum_{g\in G}( a_{g}+b_{g})g ,~~ (\sum_{g\in G} a_{g}g)(\sum_{h\in G} b_{h}h)=\sum_{g\in G, h\in G}( a_{g}b_{h})g\cdot h,
\end{equation}
where the operation '$\cdot$' denotes the multiplication of group elements. The product by a scalar is defined as
\begin{equation}
\label{eq:12}
 a(\sum_{g\in G} a_{g}g)= \sum_{g\in G} (aa_{g})g.
\end{equation}
From above definitions, we can see that a group algebra describes the superposition of symmetries expressed by group elements. Similar to the residual symmetry $Z_{2}\times CP$, the elements of a group algebra with continuous superposition coefficients may also describe partial symmetries of leptons. They may be used to predict the lepton mixing pattern. For simplicity, we consider the group algebra constructed by two group elements in this paper. Namely, the residual symmetry is expressed as
\begin{equation}
 \label{eq:13}
  X_{e,\nu}=x_{1e,\nu}A_{1e,\nu}+x_{2e,\nu}A_{2e,\nu},
\end{equation}
where $A_{1e,\nu}$, $A_{2e,\nu}$ are elements of a small group. Through equivalent transformations, the superposition coefficients are dependent on a real parameter in a special parametrization. So we can obtain clear relations between mixing parameters and the adjustable coefficient. In spite of the economy of the structure, $X_{e,\nu}$ seems strange. It is not a group element in general. The choice of $A_{i}$ seems random. To realize the characteristic of the novel construct, we study a minimal case with the $S_{3}$ group algebra. We find that $X$ in the  $S_{3}$ group algebra is equivalent to the symmetry $Z_{2}\times CP$ in the case of Dirac neutrinos. Furthermore, the maximal or trivial Dirac $CP$ phase could be obtained from $X$ in the $S_{4}$ group algebra. Although we cannot prove that the equivalence holds for $X$ in a general algebra, we may have more choices in the realization of  partial symmetries.

This paper is organised as follows. In Sec.~II, we show an economical realization of group algebras. In Sec.~III, we study a minimal case with a $S_{3}$ group algebra. Finally, we give a conclusion.

\section{Realization of a group algebra}
An element of a group algebra is constructed by the superposition of elements of a group. Here we consider the elements of group algebras obtained from two group elements. We note that the representation matrix of $X$ is not unitary in general even if the representation of the group elements is unitary. In order to obtain keep the representation of $X$ unitary, we set extra constraints on coefficients and group elements, namely
\begin{equation}
\label{eq:14}
\left \{{
\begin{array}{c}
  (|x_{1}|^{2}+|x_{2}|^{2})I+x_{1}x_{2}^{\ast}A_{1}A_{2}^{+}+x_{2}x_{1}^{\ast}A_{2}A_{1}^{+}=I,\\
(|x_{1}|^{2}+|x_{2}|^{2})I+x_{2}x_{1}^{\ast}A_{1}^{+}A_{2}+x_{1}x_{2}^{\ast}A_{2}^{+}A_{1}=I,
\end{array}}
\right.
\end{equation}
where the signal "$\ast$" denotes the complex conjugation. An economical solution to the constraints equations is
\begin{equation}
\label{eq:15}
\left \{{
\begin{array}{c}
  |x_{1}|^{2}+|x_{2}|^{2}=1,\\
e^{i\alpha}A_{1}A_{2}^{+}+e^{-i\alpha}A_{2}A_{1}^{+}=O,\\
 e^{-i\alpha}A_{1}^{+}A_{2}+e^{i\alpha}A_{2}^{+}A_{1}=O,
\end{array}}
\right.
\end{equation}
where $\alpha$ is the phase of the term $x_{1}x_{2}^{\ast}$, $O$ is the zero matrix. Up to a global phase, by a redefinition of the matrix $A_{1}$ or $A_{2}$, $X$ can be parameterized as\cite{46}
\begin{equation}
\label{eq:16}
X(\theta)=\cos\theta A_{1}+i\sin\theta A_{2},
\end{equation}
where $i$ is the imaginary factor, $A_{1}$, $A_{2}$ satisfy the constraints
\begin{equation}
\label{eq:17}
A_{1}A_{2}^{+}=A_{2}A_{1}^{+},~~ A_{1}^{+}A_{2}=A_{2}^{+}A_{1}.
\end{equation}
So $A_{1}A_{2}^{+}$, $A_{1}^{+}A_{2}$ are generators of $Z_{2}$ groups. $X$ can be rewritten as $X= A_{1}e^{i\theta B}$ with $B=A_{1}^{+}A_{2}$, $B^{2}=I$.

Let's make some necessary comments here:\\
{\bf a.} For Majorana neutrinos, the residual symmetry is $Z_{2}\times Z_{2}$. It can be broken to the partial symmetry $Z_{2}$. $X$  depends on a continuous parameter $\theta$. It is not a  $Z_{2}$ symmetry in general. So $X$  is used for the description of residual symmetries of charged leptons and Dirac neutrinos.\\
{\bf b.} With a special choice of group elements $A_{i}$ and the parameter $\theta$, $X$ could become a generator of a large cyclic group. An example is given in Ref.~\cite{46}.\\
{\bf c.} The mixing matrix from $X(\theta)$ is dependent on a parameter $\theta$. Furthermore, $X(\theta)$ is equivalent to $Z_{2}\times CP$ in the case of $S_{3}$ group algebras. This interesting observation still holds for some elements of  $S_{4}$ group algebras.\\
{\bf d.} Although $X$ is dependent on the parameter $\theta$, some mixing angle or CP phase may be independent of $\theta$. We may separate impacts of discrete group elements and $\theta$ in special cases.

\section{A minimal case for $S_{3}$ group algebra}
For illustration, we consider a minimal case that the group algebra is constructed by elements of the group $S_{3}$. Although the 3-dimensional representation of $S_{3}$ group algebras is reducible, it can be viewed as the special case of $S_{4}$ group algebras. In this section we first consider the special case that the mass matrix of charged leptons is diagonal. So the lepton mixing matrix is just dependent on the residual symmetry $X_{\nu}$. Then we show equivalence of elements of $S_{3}$ group algebras and the residual symmetry $Z_{2}\times CP$. Comments on $S_{4}$ group algebras are also made. Finally, we discuss general residual symmetries of the charged lepton sector.

\subsection{Mixing patterns from $S_{3}$ group algebra in the case of the diagonal mass matrix $M_{e}^{+}M_{e}$}
\begin{figure}[tbp]
\centering % \begin{center}/\end{center} takes some additional vertical space
\includegraphics[width=.45\textwidth]{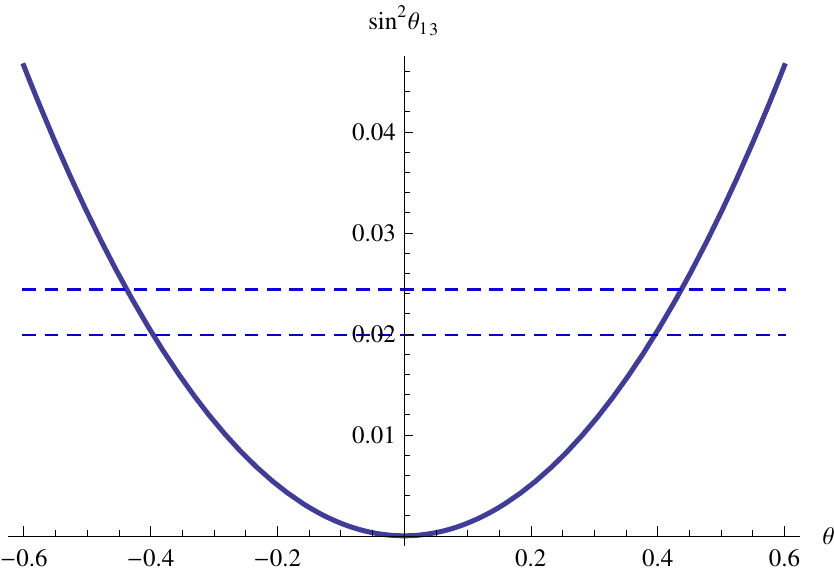}
\hfill
\includegraphics[width=.45\textwidth]{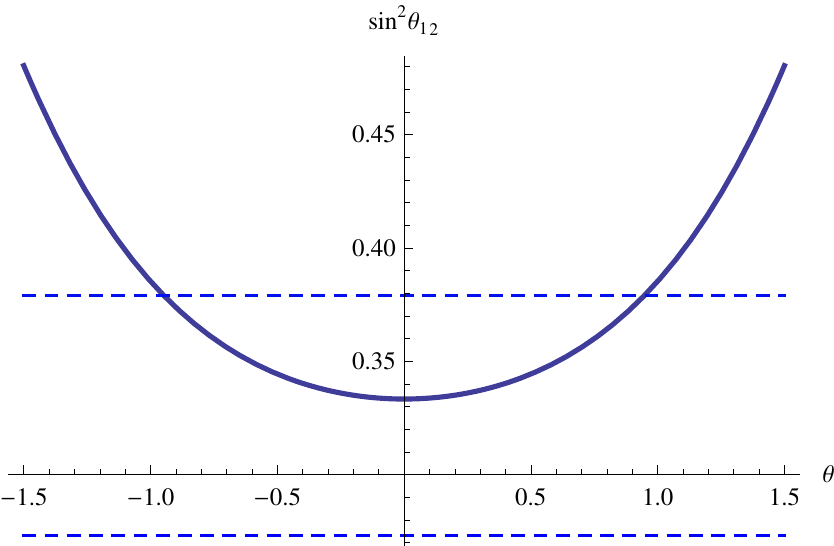}
\caption{\label{fig:1} Dependence of functions $\sin^{2}\theta_{13}$, $\sin^{2}\theta_{12}$ on the variable $\theta$. Two dashed lines in each panel label the $3\sigma$ range of the mixing angle from the recent global fit\cite{51}. For $\sin^{2}\theta_{13}$, we take the $3\sigma$ range in the normal mass ordering.  }
\end{figure}

The 3-dimensional reducible representation of the group $S_{3}$  is expressed as
\begin{equation}
\label{eq:18}
\begin{array}{c}
  \ I=\left(
                 \begin{array}{ccc}
                   1 & 0 & 0 \\
                   0 & 1 &0 \\
                   0 & 0 & 1 \\
                 \end{array}
               \right),\ S_{12}=\left(
                 \begin{array}{ccc}
                   0 & 1 & 0 \\
                   1 & 0 & 0 \\
                   0 &0 & 1\\
                 \end{array}
               \right),\ S_{13}=\left(
                 \begin{array}{ccc}
                  0 &0 & 1 \\
                   0 & 1 &0 \\
                   1 & 0 & 0 \\
                 \end{array}
               \right),\\
    S_{23}=\left(
         \begin{array}{ccc}
           1 & 0 & 0 \\
           0 & 0 & 1 \\
           0 & 1 & 0 \\
         \end{array}
       \right), ~~S_{123}=\left(
         \begin{array}{ccc}
           0 & 0 & 1 \\
           1 & 0 & 0 \\
           0 & 1 & 0 \\
         \end{array}
       \right),~~S_{132}=\left(
         \begin{array}{ccc}
           0 &1 & 0 \\
           0 & 0 & 1 \\
           1 & 0 & 0 \\
         \end{array}
       \right).
\end{array}
\end{equation}
According to the unitary conditions Eq.\ref{eq:17}, viable nontrivial realizations of $X_{\nu}$ are listed as
\begin{equation}
\label{eq:19}\begin{array}{c}
 X_{1\nu}\equiv S_{23}e^{i\theta S_{12}},~~ X_{2\nu}\equiv S_{23}e^{i\theta S_{13}}, ~~ X_{3\nu}\equiv S_{12}e^{i\theta S_{13}}, ~~X_{4\nu}\equiv S_{12}e^{i\theta S_{23}},\\
 X_{5\nu}\equiv S_{13}e^{i\theta S_{12}}, ~~X_{6\nu}\equiv S_{13}e^{i\theta S_{23}}, ~~ X_{7\nu}\equiv S_{123}e^{i\theta S_{12}}, ~~X_{8\nu}\equiv S_{123}e^{i\theta S_{23}},\\
 X_{9\nu}\equiv S_{123}e^{i\theta S_{13}}, ~~X_{10\nu}\equiv S_{132}e^{i\theta S_{12}}, ~~ X_{11\nu}\equiv S_{132}e^{i\theta S_{23}}, ~~X_{12\nu}\equiv S_{132}e^{i\theta S_{13}}.
 \end{array}
\end{equation}
All theses $X_{\nu}$  correspond to the same lepton mixing matrix up to permutations of rows, columns, or trivial phases. We consider $X_{1\nu}$ as a representative, whose expression is
\begin{equation}
\label{eq:20}
X_{1\nu}\equiv
\left(
         \begin{array}{ccc}
           \cos\theta & i\sin\theta &0 \\
           0 & 0 & e^{i\theta} \\
           i\sin\theta &\cos\theta &0  \\
         \end{array}
       \right).
\end{equation}
It is diagonalized as
\begin{equation}
\label{eq:21}
U_{\nu}^{+}X_{1\nu}U_{\nu}=diag(e^{i\theta_{1}}, ~~e^{i\theta},~~e^{i\theta_{2}}),
\end{equation}
where $e^{i\theta_{1}}\equiv\sqrt{1-s^{2}/4}-\frac{is}{2}$, $e^{i\theta_{2}}\equiv-\sqrt{1-s^{2}/4}-\frac{is}{2}$, $ s\equiv\sin\theta$. The matrix $U_{\nu}$ reads
\begin{equation}
\label{eq:22}
U_{\nu}=
\left(
         \begin{array}{ccc}
          \frac{e^{i\theta_{1}}-ce^{i(\theta-\theta_{1})}}{is\sqrt{N_{1}}} & \frac{1}{\sqrt{3}} & \frac{e^{i\theta_{2}}-ce^{i(\theta-\theta_{2})}}{is\sqrt{N_{2}}} \\
          \frac{1}{\sqrt{N_{1}}} &\frac{1}{\sqrt{3}} & \frac{1}{\sqrt{N_{2}}} \\
           \frac{e^{i(\theta-\theta_{1})}}{\sqrt{N_{1}}} & \frac{1}{\sqrt{3}} & \frac{e^{i(\theta-\theta_{2})}}{\sqrt{N_{2}}} \\
         \end{array}
       \right),
\end{equation}
where $c\equiv\cos\theta$, $N_{j}\equiv2+\frac{1+c^{2}-2c\cos(\theta-2\theta_{j})}{s^{2}}$, $j=1, 2$. It is of trimaximal form with the $\mu-\tau$ reflection symmetry\cite{47,48,49,50}, i.e., $U_{\alpha2}=\frac{1}{\sqrt{3}}$ with $\alpha=e, \mu, \tau$, $|U_{\mu j}|=|U_{\tau j}|$ with $j=1, 2, 3$. The lepton mixing matrix $U_{PMNS}$ is equal to $U_{\nu}$ up to permutations of rows or columns. Given the recent global fit data of neutrino oscillations\cite{51}, viable mixing matrices are
\begin{equation}
\label{eq:23}
U_{1}\equiv U_{\nu},~~U_{2}\equiv S_{23}U_{1}, ~~U_{3}\equiv U_{1}S_{13},~~U_{4}\equiv U_{2}S_{13}.
\end{equation}
Note that $U_{3}(\theta)=U_{1}(\theta+\pi)$, $U_{4}(\theta)=U_{2}(\theta+\pi)$. Furthermore, according to the standard parametrisation\cite{52}
 \begin{equation}
\label{eq:24}
 U_{PMNS}=\left( \begin{array}{ccc}{c_{12} c_{13}} & {s_{12} c_{13}} & {s_{13} e^{-i \delta_{CP}}} \\ {-s_{12} c_{23}-c_{12} s_{13} s_{23} e^{i \delta_{CP}}} & {c_{12} c_{23}-s_{12} s_{13} s_{23} e^{i \delta_{CP}}} & {c_{13} s_{23}} \\ {s_{12} s_{23}-c_{12} s_{13} c_{23} e^{i \delta_{CP}}} & {-c_{12} s_{23}-s_{12} s_{13} c_{23} e^{i \delta_{CP}}} & {c_{13} c_{23}}\end{array}\right)\left( \begin{array}{ccc}{e^{i \alpha_{1}}} & {0} & {0} \\ {0} & {e^{i \alpha_{2}}} & {0} \\ {0} & {0} & {1}\end{array}\right),
 \end{equation}
 where $s_{i j} \equiv \sin \theta_{i j}, c_{i j} \equiv \cos \theta_{i j}$, $\delta_{CP}$ is the Dirac CP-violating phase, $\alpha_{1}$ and $\alpha_{2}$ are Majorana phases, $U_{1}$, $U_{2}$ are interchanged through the transformation:
$\theta_{23} \rightarrow \frac{\pi}{2}-\theta_{23}$, $\delta_{CP} \rightarrow \delta_{CP}+\pi$. So without loss of generality, we can just consider $U_{1}$. Lepton mixing angles and Dirac CP phase are listed as
\begin{equation}
\label{eq:25}
\sin^{2}\theta_{13}=\frac{1+c^{2}-2c\cos(\theta-2\theta_{2})}{2+s^{2}-2c\cos(\theta-2\theta_{2})},~~\sin^{2}\theta_{12}=\frac{1}{3\cos^{2}\theta_{13}},~~\sin^{2}\theta_{23}=\frac{1}{2},
~~\delta_{CP}=-sign(s)\frac{\pi}{2},
\end{equation}
where $s\neq0$. Dependence of $\sin^{2}\theta_{13}$ and $\sin^{2}\theta_{12}$ on the variable $\theta$ is shown in Fig. \ref{fig:1}.
From the figure, we can see that $\sin^{2}\theta_{12}$ is a slowly varying function of the parameter $\theta$. So the parameter space of $\theta$ is mainly constrained by $\sin^{2}\theta_{13}$. According to the function $\chi^{2}$ defined as
\begin{equation}
\label{eq:26}
\chi^{2}=\sum_{ij=13,23,12}(\frac{\sin^{2}\theta_{ij}-(\sin^{2}\theta_{ij})_{exp}}{\sigma_{ij}})^{2},
\end{equation}
where $(\sin^{2}\theta_{ij})_{exp}$ are best global fit values from Ref.~\cite{51}, $\sigma_{ij}$ are 1$\sigma$ uncertainties,
 best fit data of  $\theta$ , $\sin^{2}{\theta_{ij}}$, $\delta_{CP}$ are listed in Table~\ref{tab:1}. They are in the $3\sigma$ ranges of the global fit data.

\begin{table}
\caption{Best fit data of the parameter $\theta$ , $\sin{\theta_{ij}}$, and $\delta_{CP}$ }
\label{tab:1}       % Give a unique label
% For LaTeX tables use
\begin{tabular}{|c|c|c|c|c|c|c|}
\noalign{\smallskip}\hline
 ~~$Ordering$~~&~~~~~$\chi^{2}_{min}~~~~~$ &~~~~~$\theta_{bf}$~~~~~&~~$(\sin^{2}\theta_{13})_{bf}$~~& ~~$(\sin^{2}\theta_{23})_{bf}$~~&~~$(\sin^{2}\theta_{12})_{bf}$~~&~~~$(\delta_{CP})_{bf}$~~ \\[0.5ex]\hline
\noalign{\smallskip}\noalign{\smallskip}\hline
 ~~ Normal~~ &~4.856~&~~$\pm0.131\pi$~~& ~0.0216~&~~0.5~~&~~0.341~&~$\mp\pi/2$\\
\hline
~~Inverted~~&~~5.855~~&~~$\pm0.132\pi$~~~&~~0.0220~~& ~0.5~&~~0.341~&~$\mp\pi/2$\\
\hline
\noalign{\smallskip}
\end{tabular}
% Or use
\vspace*{0.5cm}  % with the correct table height
\end{table}

\subsection{Equivalence of elements of $S_{3}$ group algebras and $Z_{2}\times CP$}
The neutrino mass matrix $M_{\nu}^{+}M_{\nu}$ which is invariant under the action of $X_{1\nu}$ is of the form
\begin{equation}
\label{eq:27}
M_{\nu}^{+}M_{\nu}=\left(
                     \begin{array}{ccc}
                       m_{ee} &  m_{e\mu} &  m_{e\mu}^{\ast} \\
                      m_{e\mu}^{\ast} & m_{\tau\tau} &  m_{e\mu}+m_{ee}-m_{\tau\tau} \\
                     m_{e\mu} & m_{e\mu}^{\ast}+m_{ee}-m_{\tau\tau} &  m_{\tau\tau} \\
                     \end{array}
                   \right),
\end{equation}
Where $m_{ee}$, $m_{\tau\tau}$ are real, Im$(m_{e\mu})=\frac{1}{2}( m_{ee}-m_{\tau\tau})\tan\theta $. Obviously, $M_{\nu}^{+}M_{\nu}$
follows the residual symmetry $Z_{2}\times CP$, i.e.,
\begin{equation}
\label{eq:28}
C_{Magic}^{+}(M_{\nu}^{+}M_{\nu})C_{Magic}=M_{\nu}^{+}M_{\nu},~~~~~S_{23}^{+}(M_{\nu}^{+}M_{\nu})S_{23}=(M_{\nu}^{+}M_{\nu})^{\ast},
\end{equation}
 where
\begin{equation}
\label{eq:29}
C_{Magic}\equiv\frac{1}{3}\left(
            \begin{array}{ccc}
              1 & -2 & -2 \\
              -2 & 1 & -2 \\
              -2 & -2 & 1 \\
            \end{array}
          \right), ~~~C_{Magic}^{2}=I.
\end{equation}
Correspondingly, for $X_{1\nu}$ we have
\begin{equation}
\label{eq:30}
C_{Magic}^{+}X_{1\nu}C_{Magic}=X_{1\nu},~~~~~S_{23}^{+}X_{1\nu}S_{23}=X_{2\nu}.
\end{equation}
$S_{23}$ works as the GCP for the mass matrix $M_{\nu}^{+}M_{\nu}$ on the one hand. On the other hand, it acts as an equivalent transformation for symmetries $X_{1\nu}, X_{2\nu}$. So $X_{1\nu}$ is equivalent to the residual symmetry $Z_{2}^{Magic}\times CP$.

\subsection{Comments on equivalence of elements of $S_{4}$ group algebras and $Z_{2}\times CP$}
For $S_{4}$ group with the GCP, the residual symmetries $Z_{2}\times CP$ could bring maximal or trivial Dirac $CP$
phase. We have seen that $X_{\nu}\cong Z_{2}\times CP$ in $S_{3}$ group algebras gives a maximal $CP$ phase. In fact, the equivalence can still hold for some $X$ in $S_{4}$ group algebras which are not elements of $S_{3}$  group algebras. The trivial $CP$ phase could be obtained from $X$. Here we give an example of $X$ from $S_{4}$ group algebras with a different representation. Three generators of $S_{4}$ which satisfy the relation\cite{32}
\begin{equation}
\label{eq:31}
S^{2}=V^{2}=(SV)^{2}=(TV)^{2}=I,~~
T^{3}=(ST)^{3}=I, ~~(STV)^{4}=I,
\end{equation}
are expressed as~\cite{32}
\begin{equation}
\label{eq:32}
\ S=\frac{1}{3}\left(
                 \begin{array}{ccc}
                   -1 & 2 & 2 \\
                   2 & -1 & 2 \\
                   2 & 2 & -1 \\
                 \end{array}
               \right),\ T=\left(
                 \begin{array}{ccc}
                   1 & 0 & 0 \\
                   0 & \omega^{2} & 0 \\
                   0 & 0 & \omega\\
                 \end{array}
               \right),\  V=\mp\left(
                 \begin{array}{ccc}
                   1 &0 & 0 \\
                   0 & 0 & 1 \\
                   0 & 1 & 0 \\
                 \end{array}
               \right),
\end{equation}
where $\omega=e^{i2\pi/3}$. A nontrivial example of $S_{4}$ group algebra element could be $X=(TV)\cos\theta+i\sin\theta (STV)$. Its
specific expression is of the form \cite{46}
\begin{equation}
\label{eq:33}
X(\theta)=\left(
\begin{array}{ccc}
 \frac{i}{3} \sin \theta -\cos \theta & \frac{2}{3} e^{i \pi /6} \sin \theta  & -\frac{2}{3}e^{-i\pi/6} \sin \theta  \\
 -\frac{2i}{3} \sin \theta  & \frac{2}{3}e^{i \pi /6} \sin \theta & \frac{1}{3} e^{-i\pi/6}\sin \theta -e^{-i2\pi/3}  \cos \theta  \\
 -\frac{2i}{3} \sin \theta  & -\frac{1}{3} e^{i\pi/6}\sin \theta-e^{i 2\pi/3}\cos \theta  & -\frac{2}{3}e^{-i \pi /6} \sin \theta \\
\end{array}
\right).
\end{equation}
If we take $X_{\nu}=X(\theta)$ and suppose that the mass matrix of charged leptons is diagonal, we can obtain the lepton mixing matrix written as
\begin{equation}
\label{eq:34}
U=diag(1, \omega, \omega^{2})\cdot
\left(
         \begin{array}{ccc}
          -\sqrt{\frac{2}{3}}c_{1} & \frac{1}{\sqrt{3}} &  -\sqrt{\frac{2}{3}}s_{1} \\
           \sqrt{\frac{1}{6}}c_{1}+ \sqrt{\frac{1}{2}}s_{1} &\frac{1}{\sqrt{3}} & \sqrt{\frac{1}{6}}s_{1}-\sqrt{\frac{1}{2}}c_{1} \\
            \sqrt{\frac{1}{6}}c_{1}- \sqrt{\frac{1}{2}}s_{1} & \frac{1}{\sqrt{3}} & \sqrt{\frac{1}{6}}s_{1}+\sqrt{\frac{1}{2}}c_{1} \\
         \end{array}
       \right),
\end{equation}
where $c_{1}\equiv\cos\theta_{1}$, $s_{1}\equiv\sin\theta_{1}$, $\theta_{1}$ is a parameter constrained by the mixing angle $\theta_{13}$. So the mixing pattern is of trimaximal form with a trivial Dirac $CP$ violating phase. For $X(\theta)$, we can verify that the following relation holds,  i.e.,
\begin{equation}
\label{eq:35}
C_{1}^{+}X(\theta)C_{1}=X(\theta),
\end{equation}
where $C_{1}=T^{+}ST$, $C_{1}^{2}=I$, $T^{+}C_{1}T=C_{1}^{\ast}$. So $C_{1}$ and $T$ are a $Z_{2}$ symmetry and the corresponding $CP$ transformation respectively. Following the methods used in GCP\cite{54},
the lepton mixing matrix from the residual symmetry $Z_{2}\times CP$ can be expressed as $U_{a}=\Omega R_{13}(\theta_{1})P$, where $\Omega$, $R_{13}(\theta_{1})$ are expressed respectively as
\begin{equation}
\label{eq:36}
 \Omega=
\left(
         \begin{array}{ccc}
          -\sqrt{\frac{2}{3}} & \frac{1}{\sqrt{3}} &  0\\
           \frac{\omega}{\sqrt{6}}&\frac{\omega}{\sqrt{3}} & \frac{-\omega}{\sqrt{2}} \\
           \frac{\omega^{2}}{\sqrt{6}} & \frac{\omega^{2}}{\sqrt{3}} & \frac{\omega^{2}}{\sqrt{2}} \\
         \end{array}
       \right),
~~R_{13}(\theta_{1})=\left(
                          \begin{array}{ccc}
                            \cos\theta_{1} & 0 &\sin\theta_{1} \\
                            0& 1 & 0 \\
                           -\sin\theta_{1} & 0& \cos\theta_{1} \\
                          \end{array}
                        \right),
\end{equation}
$P$ is a phase matrix which can be neglected in our case of Dirac neutrinos. Specially,
the matrix $\Omega$ satisfies the relations as follows
\begin{equation}
\label{eq:37}
 \Omega^{+}C_{1}\Omega=diag(-1,~1,~-1), ~~ T=\Omega\Omega^{T}.
\end{equation}
We can check that the matrix  $U_{a}$ from the $Z_{2}\times CP$ is just the $U$ shown in Eq.~\ref{eq:34}. So $X(\theta)$ is equivalent to the symmetry $Z_{2}\times CP$ generated by  $C_{1}$ and $T$. Furthermore, let's consider the element $ X^{\prime}(\theta)\equiv T^{+}X(\theta)T$. The lepton mixing matrix from $ X^{\prime}(\theta)$ is $U^{\prime}=T^{+}U$. Since $T$ is a phase matrix, $U^{\prime}$ is equivalent to $U$. So the $CP$ transformation
interchanges the equivalent elements  $X(\theta)$, $ X^{\prime}(\theta)$. Therefore, the observation from the case of the $S_{3}$  algebra still holds in this example of the $S_{4}$ group algebra.

\subsection{Discussion on general residual symmetries of the charged lepton sector }
We have studied the case that the mass matrix $M^{+}_{e}M_{e}$ is diagonal. The corresponding symmetry of the charged lepton sector is $U(1)\times U(1)\times U(1)$, namely $X_{e}=diag(e^{i\alpha_{1}},~e^{i\alpha_{2}},~e^{i\alpha_{3}})$. Now we discuss a more general case that $X_{e}$ is expressed by an element of the $S_{3}$ group algebra. Because all the elements listed in Eq.~\ref{eq:19} give the same mixing matrix up to permutations of rows or columns, we can take  $X_{1e}= S_{23}e^{i\theta_{e} S_{12}}$. Then the matrix $U_{e}$ is of the form
\begin{equation}
\label{eq:38}
U_{e}=
\left(
         \begin{array}{ccc}
          \frac{e^{i\theta_{1e}}-c_{e}e^{i(\theta_{e}-\theta_{1e})}}{is_{e}\sqrt{N_{1e}}} & \frac{1}{\sqrt{3}} & \frac{e^{i\theta_{2e}}-c_{e}e^{i(\theta_{e}-\theta_{2e})}}{is_{e}\sqrt{N_{2e}}} \\
          \frac{1}{\sqrt{N_{1e}}} &\frac{1}{\sqrt{3}} & \frac{1}{\sqrt{N_{2e}}} \\
           \frac{e^{i(\theta_{e}-\theta_{1e})}}{\sqrt{N_{1e}}} & \frac{1}{\sqrt{3}} & \frac{e^{i(\theta_{e}-\theta_{2e})}}{\sqrt{N_{2e}}} \\
         \end{array}
       \right),
\end{equation}
where $e^{i\theta_{1e}}\equiv\sqrt{1-s_{e}^{2}/4}-\frac{is_{e}}{2}$, $e^{i\theta_{2e}}\equiv-\sqrt{1-s_{e}^{2}/4}-\frac{is_{e}}{2}$, $ s_{e}\equiv\sin\theta_{e}$, $c_{e}\equiv\cos\theta_{e}$, $N_{je}\equiv2+\frac{1+c_{e}^{2}-2c_{e}\cos(\theta_{e}-2\theta_{je})}{s_{e}^{2}}, j=1,~2$.  With respect to the mixing matrix $U_{PMNS}\equiv U_{e}^{+}U_{\nu}$, we have an element $U_{PMNS}(\alpha i)=1$. Obviously,
it does not satisfy the constraint of the global fit data of neutrino oscillations. So the combination of the residual symmetries ($X_{1e},~X_{1\nu}$) does not give a realistic lepton mixing patten in the case of $S_{3}$ group algebra. Furthermore, if $\theta_{e}$ is equal to 0, $X_{1e}$ is reduced to $S_{23}$. The corresponding matrix $U_{e}$ becomes
\begin{equation}
\label{eq:39}
U^{\prime}_{e}=
\left(
         \begin{array}{ccc}
          0 & -\sin\theta^{\prime}&\cos\theta^{\prime}\\
         \frac{-1}{\sqrt{2}} &\frac{\cos\theta^{\prime}}{\sqrt{2}} & \frac{\sin\theta^{\prime}}{\sqrt{2}} \\
  \frac{1}{\sqrt{2}} & \frac{\cos\theta^{\prime}}{\sqrt{2}} &  \frac{\sin\theta^{\prime}}{\sqrt{2}}\\
         \end{array}
       \right),
\end{equation}
where $\theta^{\prime}$ is an angle variable from the degeneracy of the eigenvalues of $S_{23}$. Then $U_{PMNS}$ contains a zero element. This observation still holds when $S_{23}$ is replaced by $S_{12}$ or $S_{13}$.
So the combination ($Z_{2e}, ~X_{1\nu}$) is not a viable choice for the residual symmetries of leptons. We can also check that $U_{PMNS}$ from the combination ($Z_{3e}, ~X_{1\nu}$), where $Z_{3e}$ is generated by $S_{123}$ or $S_{132}$, does not satisfy the constraint of the global fit data of neutrino oscillations either. It contains an element which is equal to 1. Therefore, when the residual symmetry of neutrinos sector is $X_{1\nu}$ in the $S_{3}$ group algebra, we can only take $X_{e}=diag(e^{i\alpha_{1}},~e^{i\alpha_{2}},~e^{i\alpha_{3}})$.

\section{Conclusion}
We have studied a new structure to describe partial symmetries of charged leptons and Dirac neutrinos. The residual symmetry is expressed by an element of group algebras. In our construction, a specific lepton mixing pattern corresponds to a set of equivalent residual symmetries which are expressed by elements of group algebras $X_{i}$. These equivalent symmetries $X_{i}$ can be interchanged through a transformation which corresponds to a residual $CP$ symmetry. For $S_{3}$ group algebras and a special case of $S_{4}$ group algebras, we found that $X_{i}$ is equivalent to a residual symmetry $Z_{2}\times CP$. The corresponding lepton mixing matrix is trimaximal. It is a difficult mathematical problem for us to determine whether $X_{i}$ is equivalent to $Z_{2}\times CP$ in general cases. Even so, observations from simple examples could still give us some interesting clues: {\bf a.} The parameter in partial symmetries may be viewed as a quantity to measure how discrete symmetries are mixed in the residual symmetry. {\bf b.}  A partial symmetry dependent on a continuous parameter may be equivalent to a discrete symmetry with GCP. ~{\bf c.} The elementary residual $CP$ transformation could be a permutation matrix or a diagonal phase matrix. A general one may be a finite product of elementary ones. Therefore, despite of stringent experiment data, we could still construct some novel partial symmetries to obtain viable lepton mixing patterns.

\acknowledgments
 This work is supported by the National Natural Science Foundation of China under grant No. 11405101, 11705113, the Guangxi Scientific Programm Foundation under grant No. Guike AD19110045, the Research Foundation of Gunlin University of Technology under grant No. GUTQDJJ2018103.\\

{\bf Competing Interests}\\
The author declares that there is no conflict of interest regarding the publication of this paper.

\end{document}